\documentclass[submission,copyright,creativecommons]{eptcs}
\providecommand{\event}{QSQW19} 
\usepackage{breakurl}             
\usepackage{underscore}           
\usepackage{graphicx,color}
\usepackage{braket}
\newcommand{\Ai}{\operatorname{Ai}}

\usepackage{amsmath}
\usepackage{amsfonts}

\title{Theoretical Studies \\on Quantum Walks with a Time-varying Coin}
\author{Haruna Katayama\qquad\qquad Noriyuki Hatakenaka
\institute{Graduate School of Integrated Arts and Sciences\\
Hiroshima University
\\
Hiroshima, Japan}
\email{halna496@gmail.com\qquad\qquad noriyuki@hiroshima-u.ac.jp}
\and
Toshiyuki Fujii
\institute{
Department of Physics \\
Asahikawa Medical University\\
Hokkaido, Japan}
\email{tfujii@asahikawa-med.ac.jp}
}
\def\titlerunning{Theoretical studies on quantum walks with a time-varying coin}
\def\authorrunning{H. Katayama, N. Hatakenaka \& T. Fujii}
\begin{document}
\maketitle

\begin{abstract}
Quantum walks can reconstruct quantum algorithms for quantum computation, 
where the precise controls of quantum state transfers between arbitrary distant sites are required. 
Here, we investigate quantum walks using a periodically time-varying coin both numerically and analytically, 
in order to explore the controllability of quantum walks while preserving its random nature.  
\end{abstract}

\section{Introduction}
The quantum computer promises drastically fast processing using quantum algorithms based on unique features of quantum mechanics. Just as classical random walks have been used to design computational algorithms in computer science, quantum walks\cite{Aharnov} also have been used to reconstruct algorithms suitable for quantum mechanics and play an incredibly important role in quantum information processing and computation protocols\cite{Childs2009,Lovett2010}. 

A state transfer is required in such quantum algorithms. It is, however,  difficult to control the state transfer because the quantum walk is inherently random\cite{Chen,Karski174,Yalnkaya,Zhan2014}. 
One of the propositions to solve this problem is a coin that has a part in determining the movement of the quantum walk.
 Wojcik et al. extended the coins that previously, were spatially constant to space-dependent coins\cite{wojcik2004quasiperiodic}. 
 Ba\~nuls et al.  diversified their work  further into the time domain, led to a more generalized quantum walk\cite{Banuls}.
 They found that a time-varying coin\cite{Albertini2009,Banuls,Cedzich2016,DiMolfetta2015,Panahiyan2018,Shikano2014,Xue} can modify trajectories of the quantum walks. 
 This implies that quantum walks are controllable.

In this paper, we present that the time-varying coins can control the trajectories on demand while preserving random processes. 
First, we numerically show that the trajectories can be surely modified by using the simplest time-varying coin, i.e., linear time dependent coin. 
Then, we discuss on the walking mechanism using the analytic solution derived by means of the long-wavelength approximation method by Knight, Rold\'{a}n, and Sipe\cite{Knight}. Moreover, we investigate quantum walks with another important coin which is more general time-dependent in both  numerical and analytical perspective, leading to the controllability of the quantum walks by the time-varying coins. 
Controlling the quantum walks can improve the quantum algorithm to be more efficient.

\section{Quantum walk with a time-varying coin}
Let us consider a quantum walk with {\it a time-varying coin} on a line. In random walks, the walker moves one step to right or left depending on the outcome of the coin toss. 
A quantum walk is a quantum-mechanical counterpart of the classical random walk. Then, a quantum walk lives in a bipartite system comprising positions and coin degrees of freedom represented as an entire Hilbert space $\mathcal{H}=\mathcal{H}_p\otimes\mathcal{H}_c$, where 
$\mathcal{H}_p$ and $\mathcal{H}_c$ are the Hilbert spaces of the walker's position and the coin spanned by $\{\ket{m}, m\in\mathbb{Z}\}$ and $\{\ket{R},\ket{L}\}$, respectively. 
The walker's behaviour is characterized by both the coin and the shift operators. The time-varying coin {\it unitary} operator employed in this paper is defined as 
\begin{align}
\hat{C}=
\begin{pmatrix}
\cos(\theta(t))& \sin(\theta(t)) \\
\sin(\theta(t))& -\cos(\theta(t)) \\
\end{pmatrix},
\label{coin}
\end{align}
which transforms the internal state of a coin through a  time-varying phase $\theta(t)$. On the other hand, the shift operator given by
\begin{align}
\hat{S}\ket{m,R}=\ket{m+1,R},\\
\hat{S}\ket{m,L}=\ket{m-1,L},
\end{align}
shifts the walker's position. 

The state after $n$ steps $\ket{\psi}_n$ is described by
\begin{align}
\ket{\psi}_n=(\hat{S}\hat{C})^n\ket{\psi}_0,
\end{align}
where $\ket{\psi}_0$ represents the walker's initial state. 
This is also re-expressed by
\begin{align}
\ket{\psi}_n&=\sum_{m=-n}^{n}[R_{m,n}\ket{m,R}+L_{m,n}\ket{m,L}],
\end{align}
if we introduce $|R(L)_{m,n}|^2$ representing the probability of being in the internal state $R(L)$ at position $m$ after $n$ steps. 
Here, the probability amplitudes $R_{m,n}$ and $L_{m,n}$ at position $m$ after $n$ steps are determined by
\begin{align}
R_{m,n+1}&=\cos\theta_n R_{m-1,n}+\sin\theta_nL_{m-1,n},\label{Rzenkasiki}\\
L_{m,n+1}&=\sin\theta_n R_{m+1,n}-\cos\theta_nL_{m+1,n},\label{Lzenkasiki}
\end{align}
where $\theta_n=\theta(nT)$ with $T$ being the time interval between steps.  Therefore, the probability of finding the walker at position $m$ after $n$ steps $P_m(n)$ is described by
\begin{align}
P_m(n)&=P_m^R(n)+P_m^L(n),\label{plobability-distribution}\\
P_m^R(n)&=|R_{m,n}|^2,\quad P_m^L(n)=|L_{m,n}|^2.
\end{align}

\section{Linear time evolution of the phase in the coin operator}
Here, let us discuss the linear time evolution of the phase, i e.,  $\theta(t)=\theta_0+\omega t$ in the coin operator as the simplest example of time-varying coins. 
Through this example, we will extract the essential properties of a quantum walk with a time-varying coin and elucidate the walking mechanism.
\subsection{Numerical approaches}
Numerical simulations are carried out by using $\theta(t)=\theta_0+\omega t$ as a specific time-dependent expression of the coin operator in Eq. \eqref{coin} with various values for the initial phase $\theta_0$ and the frequency $\omega$. 
 Figure 1 (A), (B) and (C) show the probability distributions of finding a walker on a line as a function of time under the identical frequency $\omega=\pi/60$ (upper panels) and their {\it walker's trajectories} representing the place where the existence probability is high (lower panels). It is found that the time-varying coin leads to periodic walker's trajectories classified into three types depending on the initial phase of the coin $\theta_0$: (1) {\bf Loop-line chain} which is composed of loops and lines appearing alternately as shown in Fig. 1(A), 
 (2) {\bf Crossing loop-loop chain} where two major trajectories intersect as shown Fig. 1 (B),
 and (3) {\bf Touching loop-loop chain} where two major trajectories come into contact as shown in Fig. 1 (C).  In addition, walker's trajectories are also modified by the frequency $\omega$. As the frequency $\omega$ increases, the size of the loops tends to become smaller in time and narrower in space (not displayed in Fig. 1).

\begin{figure}[htbp] 
 \centering
 \includegraphics[width=15cm]{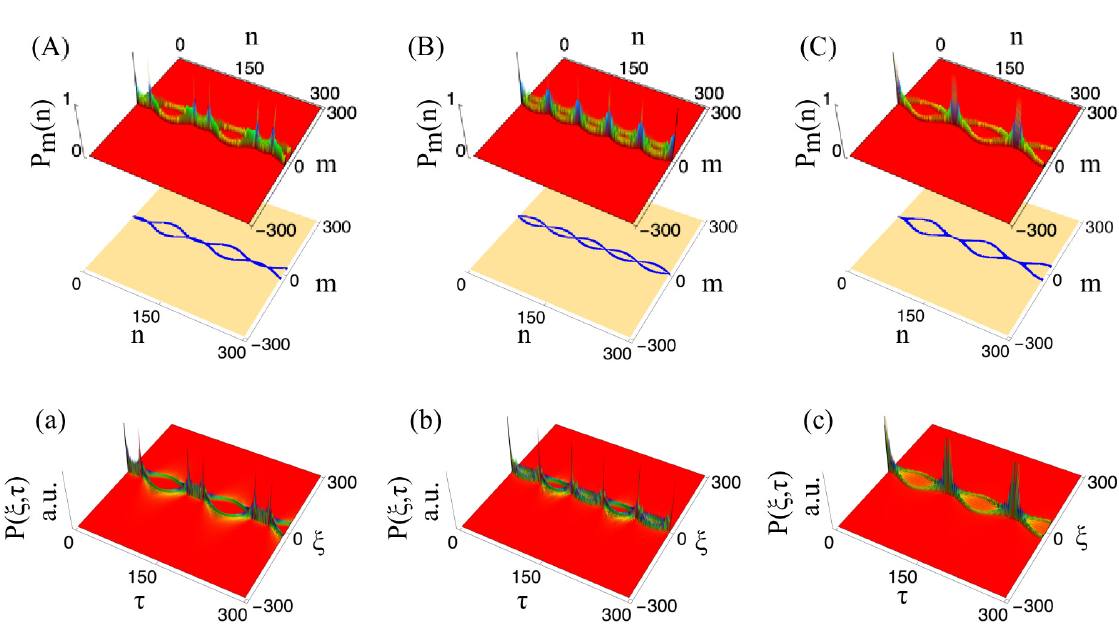}
  \caption{Probability distributions obtained by numerical simulations (upper panels) and their walker's trajectories (lower panels) for quantum walks over 300 steps with the time interval between steps $T=1$ and initial state $\ket{\psi}_0=\ket{0,R}/\sqrt{2}+i\ket{0,L}/\sqrt{2}$, for various initial phases (A)$\theta_0=\pi/4$, (B)$\theta_0=0$, and (C)$\theta_0=3\pi/2$ under  the fixed coin's frequency $\omega=\pi/60$. 
  The corresponding analytical solutions are given in (a), (b) and (c) of the bottom panels.}
\end{figure}

\subsection{Analytical approach }
In order to explore the physical origin of walker's trajectories obtained by numerical simulations in the previous section, analytical probability distributions are obtained according to the method proposed by Knight, Rold\'{a}n and Sipe\cite{Knight}. Equations (\ref{Rzenkasiki}) and (\ref{Lzenkasiki}) are reduced to the following recurrence formula,
\begin{align}
A_{m,n+1}-A_{m,n-1}&=-\cos{\theta_n}(A_{m+1,n}-A_{m-1,n}),
\label{=sabun}
\end{align}
where $\theta_n=\theta_0+n \omega T$ and $A=R,L$. In the long-wavelength approximation, we obtain the wave equation for the continuous function $A(x,t)$
\begin{align}
T\frac{\partial}{\partial t}A(x,t)=-\cos{\theta(t)}\left(X\frac{\partial}{\partial x}+\frac{X^{3}}{3!}\frac{\partial^{3}}{\partial x^{3}}\right)A(x,t).\label{bibun-mae}
\end{align}
where $X$ stands for the distance between steps.
We further introduce other two fields $A^{+}(\xi,\tau)$ and $A^{-}(\xi,\tau)$ to express the waves spreading in both sides on a line,  resulting in the wave equation
\begin{align}
\frac{\partial}{\partial \tau}A^{\pm}(\xi,\tau)=\mp\cos{\theta(\tau)}\left(\frac{\partial}{\partial \xi}+\frac{1}{3!}\frac{\partial^{3}}{\partial \xi^{3}}\right)A^{\pm}(\xi,\tau),
\label{=A+-bibun}
\end{align}
where $\xi=x/X$ and $\tau=t/T$. 
This differential equation can be solved using Fourier techniques after separation of variables in a usual manner. 
The solution for Eq. (\ref{=A+-bibun}) are given as follows
\begin{align}
A^{\pm}(\xi,\tau)&=\frac{1}{2}\sum_{m}(A_{m,0}\pm A_{m,1})Z^{\pm}(\xi-m,\tau),\label{solution1}
\end{align}
with 
\begin{align}
Z^{\pm}(\xi,\tau)&=\left|\frac{2}{s(\tau)}\right|^{\frac{1}{3}}e^{\chi}\Ai(\zeta),\label{solution2}\\
\zeta&=\left|\frac{2}{s(\tau)}\right|^{\frac{1}{3}}\left(\pm\xi-  s(\tau)+\frac{2w^4}{s(\tau)}\right),\label{solution3}\\
\chi&=\frac{2 w^2}{s(\tau)}\left(\pm \xi- s(\tau)+\frac{4 w^4}{3s(\tau)}\right),\label{solution4}\\
s(\tau)&=\mp \int_{0}^{\tau}\cos{\theta(\tau')}d\tau',\label{solution5}
\end{align}
where $\Ai$ represents the Airy function.
The probability distribution of finding a walker is then 
given by
\begin{align}
P(\xi,\tau)&=P^{R}(\xi,\tau)+P^{L}(\xi,\tau),\label{analytic-probability}\\
P^{A}(\xi,\tau)&=|A^{+}(\xi,\tau)+(-1)^{n}A^{-}(\xi,\tau)|^2.
\end{align}
Figure 1 (a), (b) and (c) show probability distributions depicted by the analytic solution for the same value as coin parameters used in  numerical simulations to (A), (B) and (C), respectively. Our analytical solution reproduces the numerical simulation well. 

One definite example is as follows. Figure 2 shows the top view of the probability distributions obtained by the analytical solution (left) and the numerical simulation (right) at $\theta_0=0$ and $\omega=\pi/60$. 
According to the probability distribution given analytically in Eq.(\ref{analytic-probability}), 
the probability $P(\xi,\tau)$ is composed of  $P^{R}(\xi,\tau)$ and $P^{L}(\xi,\tau)$. 
The highest probabilities in the analytical probability distribution are highlighted in blue ($P^{R}(\xi,\tau)$) and red ($P^{L}(\xi,\tau)$). The blue line and red line are exactly intersecting at $\tau=60$ with the same amplitude, resulting in a crossing loop-loop chain as seen in the numerical simulation. 
Actually, there exists a slight difference between analytical solutions and numerical simulations. 
Specifically, the amplitudes of the first and second loops differ only slightly in the graph on the left of Fig. 2. 
This is due to the approximation limit based on the wave nature of the quantum walk, i e., the nature of the Airy function. 
The maximum peak of the Airy function $\Ai(x)$  is not located at $x=0$, but at slightly negative position. 
As a result, the blue peak shifts more negatively than the dotted green line, and the red peak does the opposite.
In other words, the maximum peak exists at the inside of the numerical loop at $\tau=30$, and outside at $\tau=90$ because the direction of the Airy function is reversed. 
This tiny difference causes only a few deviations. Thus, the analytical solution approximates the numerical simulation well.
However, the linear spreading inherent in the quantum walk is not reproduced in
the analytical solution 
and remains unsolved.

Despite the above better agreement, the probability distribution shown in Fig. 1 (b) may look  different from that of numerical simulation shown in Fig. 1 (B). 
This difference is due to the Poggendorff illusion\cite{Suzanne},
which is a geometrical-optical illusion that involves the brain's perception of the interaction between diagonal lines and horizontal and vertical edges. 
In fact, the probability of being inward in the first (left) loop and outward in the second (right) loop causes the Poggendorff illusion. 
In contrast, the numerical simulation does not show such a difference, resulting in no Poggendorff illusion. 
From the above, the analytic solution seemed to be different from the probability distribution of the numerical simulation, but in fact the analytic solution could approximate the numerical simulation correctly.

\begin{figure}[htbp] 
 \centering
 \includegraphics[width=15cm]{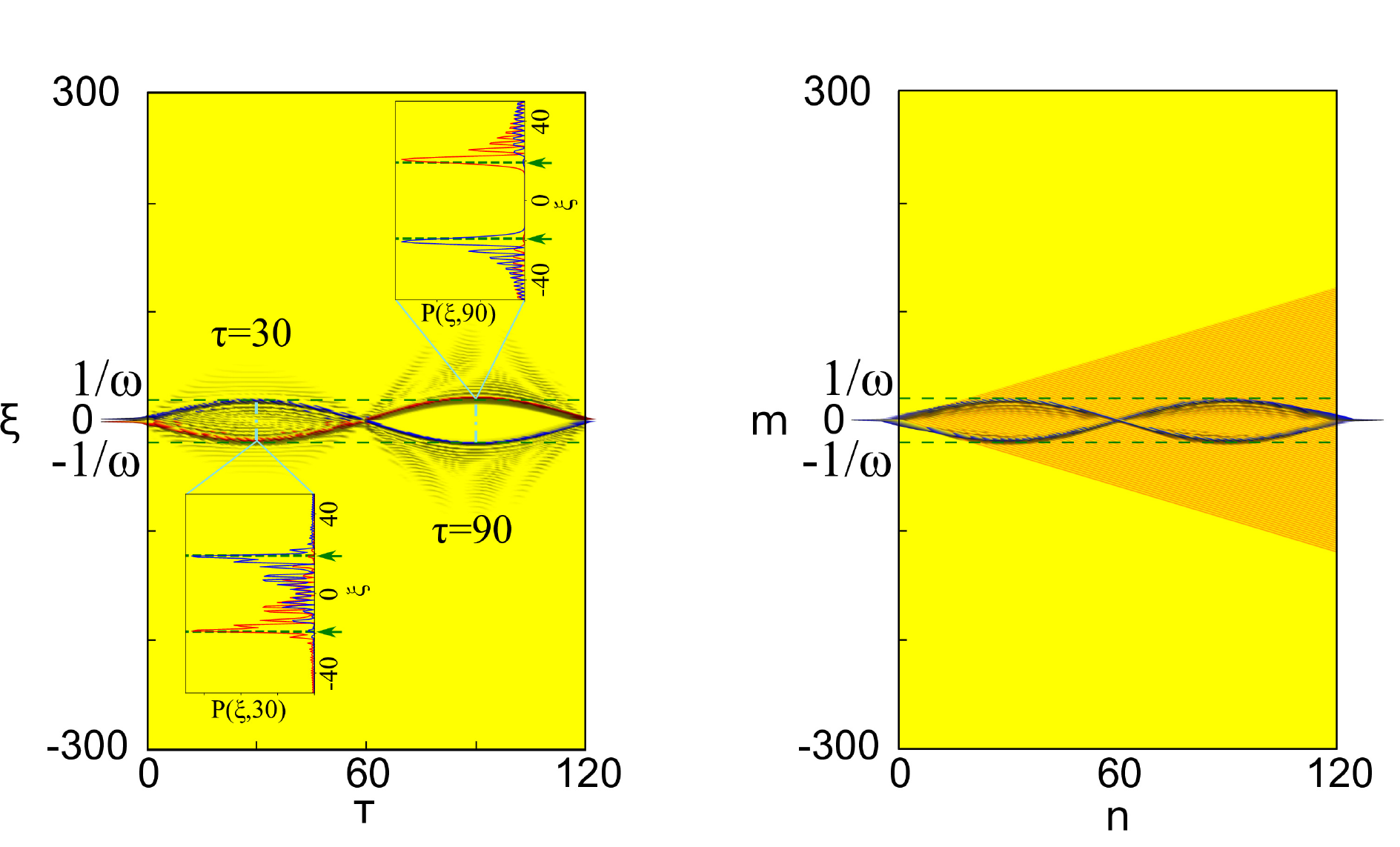}
  \caption{Top views of the probability distributions depicted in Fig.1 (B) and (b) over 120 steps. The analytical probability distributions are shown on the left,  where the high probabilities formed by $P^{R}(\xi,\tau)$ and $P^{L}(\xi,\tau)$ are highlighted in blue and red with shaded lower existence probability. Two embedded figures represent the sectional view of the probability distribution at $\tau=30$ and $\tau=90$, respectively.
    The numerical probability distribution displayed on the right is formed by a trail with high probabilities highlighted in blue and a triangular area with low probabilities highlighted in orange, which are not drawn in Fig. 1(B) due to the accuracy of the color gradation. The values of the probability distributions are equal to zero in yellow area of both diagrams. The doted lines in both diagrams represent the amplitude of the walker's trajectory given analytically in Eq. \eqref{trajectory}}
\end{figure}

\subsection{Waking mechanism}
Now, let us discuss the walking mechanism based on the analytical solutions supported by numerical simulations obtained above. 
To find the walker's trajectories which are a key clue for elucidating the walking mechanism, we examine the behavior of the probability amplitude $A^{\pm}(\xi,\tau)$, which is the main element that determines the probability distribution.
The amplitude $A^{\pm}(\xi,\tau)$ is expressed as follows:

\begin{align}
A^{\pm}(\xi,\tau)
&=\int_{-\infty}^{\infty}e^{ik\xi}A^{\pm}(k,\tau)dk \nonumber \\
&=B^{\pm}\left(\xi\mp \int_{0}^{\tau}\cos{\theta(\tau')}d\tau',\tau\right),
\label{B-shift}
\end{align}
where $B^{\pm}(\xi,\tau)$ represents a shifted probability amplitude described by
\begin{align}
B^{\pm}\left(\xi,\tau\right)
=\int_{-\infty}^{\infty}e^{ik\xi}\left(e^{\pm\frac{1}{6}ik^3\int_{0}^{\tau}\cos{\theta(\tau')}d\tau'}A^{\pm}(k,0)\right)dk.
\end{align}
Therefore, the walker's trajectories is analytically obtained by the shift in Eq. \eqref{B-shift} as follows:
\begin{align}
x_c&=\mp \int_{0}^{\tau}\cos{\theta(\tau')}d\tau' \nonumber \\
&=\mp\left(\frac{1}{\omega}\sin{(\theta_0+\omega\tau)}-\frac{1}{\omega}\sin{\theta_0}\right)\label{trajectory}.
\end{align}
Equation (\ref{trajectory}) means that walker's trajectories are composed of two antisymmetric  sinusoidal functions with same parameters such as amplitude $1/\omega$, initial phase $\theta_0$, frequency $\omega$, and bias $-\sin\theta_0/\omega$. The walker's trajectories are formed by overlap of two sinusoidal waves, and are classified into three categories like loop-line chains, crossing loop-loop chains, and touching loop-loop chains as seen in numerical simulations. 
According to our simple overlap model, these three types of chains are interpreted as follows.
Figure 3 (a) shows the first type of walker's trajectories named "loop-line chain".  In fact, two different size loops appear alternately though loop and line seemed to appear alternately. This type of chains is generated when the bias is finite except for $\theta_0=n\pi/2, (n=$integer).
Figure 3 (b) shows the second type of walker's trajectories, named  "crossing loop-loop chain". 
This type of chains occurs when the bias is  equal to 0, equivalently $\theta_0=n\pi$.
Figure 3 (c) shows the third type of walker's trajectories "touching loop-loop chain", which appears  when the sinusoidal-wave amplitude equals its bias $(\theta_0=(2n+1)\pi/2)$. In this way, 
based on our analytical solution, it is found that walker's trajectories are classified into three depending on the $\theta_0$ as obtained by numerical simulation.

The walker's trajectories reflecting the walking mechanism reveal its physical origin. 
A trajectory is, in general, simply described by $\int v(t) dt$ wiht $v(t)$ being time-dependent velocity. 
According to wave theory, the velocity $v(t)$ works to connect the wave number $k$ (space) and frequency $\omega$ (time) through the well-known dispersion relation $\omega = v k$ in the linear wave. From our wave equation, Eq.(\ref{=A+-bibun}), the term $\mp \cos{\theta(\tau)}$ corresponds to the wave velocity. In other words, the velocity that determines the trajectory is nothing but the transition amplitude of the coin itself. This shows that the time-varying coin parameter, i.e., the wave velocity, controls the walker's trajectories. 
This is our central result of this paper.

\begin{figure}[htbp]
  \centering
  \includegraphics[width=15cm]{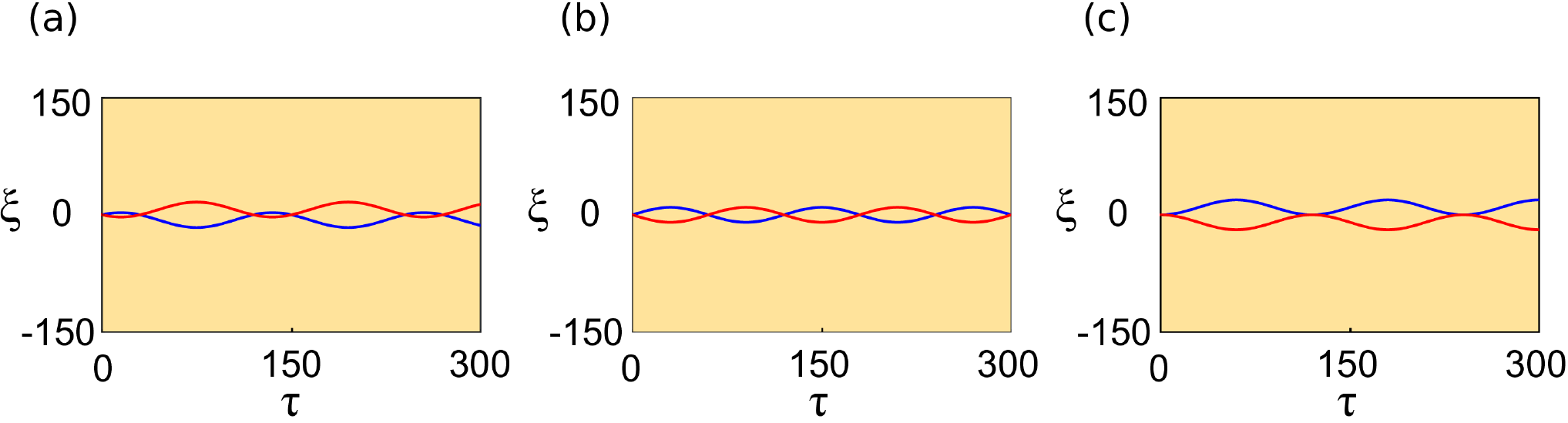}
  \caption{Three types of walker's trajectories depicted by the analytic solution for the quantum walk over 300 steps with a time-varying coins for identical coin frequency $\omega=\pi/60$ (a) a loop-line chain emerging with $\theta_0=\pi/4$ (b) a crossing loop-loop chain emerging with $\theta_0=n \pi$ (c) a touching loop-loop chain emerging with $\theta_0=(2n+1) \pi/2$}
\end{figure}

\section{Sinusoidal time evolution of the phase in the coin operator}
In the previous section, we have revealed the walking mechanism of the quantum walk with a time-varying coin for the simplest example. Here let us apply this mechanism to other time-varying coin such as $\theta(t)=\theta_0\sin(\omega t)$. 
This is considered to be a great clue for extending to coins with general time dependence because $\theta_0\sin(\omega t)$ is a component of the Fourier series expansion which describes general time dependences.

The analytic solutions Eqs. \eqref{solution1}-\eqref{solution5} are derived parallel to the previous section. Figure 4 shows the probability distribution obtained both numerically and analytically. The analytical solutions agree well with the numerical simulations. The walker's trajectory is similarly obtained  by
\begin{align}
x_c&=\mp \int_{0}^{\tau}
\cos{\left(\theta_0\sin(\omega \tau')\right)}
d\tau'\nonumber\\
&=\mp\left\{J_0(\theta_0)\tau+\sum_{k=1}^{\infty}J_{2k}(\theta_0)\sin(2k\omega\tau))\right\}\label{trajectory-sin}
\end{align}
where $J_{k}(\theta_0)$ is the Bessel function of the first kind of order $k$ of $\theta_0$.  The walker's trajectory spreads linearly owing to the first term $J_0(\theta_0)\tau$ in Eq. (\ref{trajectory-sin}), and fluctuates periodically around the linear term due to the rest of the other terms. 
In practice, the trajectory can be described accurately with only a few terms in the summation due to the Bessel functions. 
As a result, it is found that our walking mechanism can be applicable to the walker's trajectories on the quantum walk with more general time-varying coin. More general extensions will appear elsewhere.

\begin{figure}[htbp] 
 \centering
 \includegraphics[width=15cm]{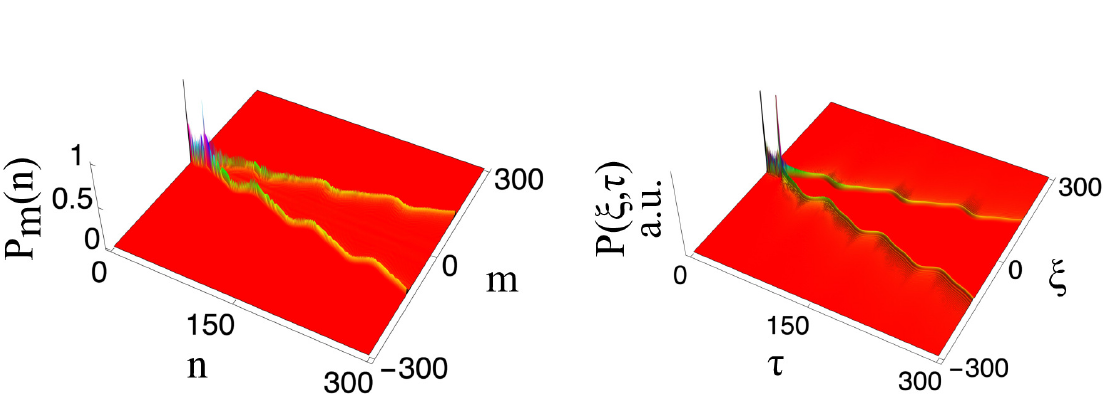}
  \caption{Probability distributions obtained numerically (left) and analytically (right) for quantum walks over 300 steps with the time-varying coin $\theta(t)=(5\pi/4)\sin(\pi t/80)$ and initial state $\ket{\psi}_0=\ket{0,R}/\sqrt{2}+i\ket{0,L}/\sqrt{2}$. 
  The other parameters are the same as in Fig. 1.
  }
\end{figure}

\section{Conclusion}
We have investigated quantum walks with a  time-varying coin, 
in order to explore the controllability of quantum walks for efficient implementations of quantum state transfers in quantum walk-based quantum computers. 
Our numerical simulations have revealed the key fundamental behaviour behind the time-varying coined quantum walk through the walker's trajectory analysis at the simplest linear time-varying coin. 
Then, an analytical solution based on the wave nature of the quantum walk has unraveled the walking mechanism that the wave velocity of the quantum walk is governed by the time-varying coin parameter. 
This mechanism has been shown to be applicable to  more important time-varying coins such as sinusoidal functions. Therefore, 
the walking mechanism enables us to control quantum walks 
by manipulating coin parameters. 
The designed time-varying coined quantum walk opens a new pathway for manipulating quantum state transfers in quantum information technology.

\nocite{*}
\bibliographystyle{eptcs}
\bibliography{generic}

\begin{thebibliography}{10}
\providecommand{\bibitemdeclare}[2]{}
\providecommand{\surnamestart}{}
\providecommand{\surnameend}{}
\providecommand{\urlprefix}{Available at }
\providecommand{\url}[1]{\texttt{#1}}
\providecommand{\href}[2]{\texttt{#2}}
\providecommand{\urlalt}[2]{\href{#1}{#2}}
\providecommand{\doi}[1]{doi:\urlalt{http://dx.doi.org/#1}{#1}}
\providecommand{\bibinfo}[2]{#2}

\bibitemdeclare{article}{Aharnov}
\bibitem{Aharnov}
\bibinfo{author}{Y.~\surnamestart Aharonov\surnameend},
  \bibinfo{author}{L.~\surnamestart Davidovich\surnameend} \&
  \bibinfo{author}{N.~\surnamestart Zagury\surnameend} (\bibinfo{year}{1993}):
  \emph{\bibinfo{title}{Quantum random walks}}.
\newblock {\sl \bibinfo{journal}{Phys. Rev. A}} \bibinfo{volume}{48}, pp.
  \bibinfo{pages}{1687--1690}, \doi{10.1103/PhysRevA.48.1687}.

\bibitemdeclare{article}{Albertini2009}
\bibitem{Albertini2009}
\bibinfo{author}{Francesca \surnamestart Albertini\surnameend} \&
  \bibinfo{author}{Domenico \surnamestart D’Alessandro\surnameend}
  (\bibinfo{year}{2009}): \emph{\bibinfo{title}{Analysis of quantum walks with
  time-varying coin on d-dimensional lattices}}.
\newblock {\sl \bibinfo{journal}{Journal of Mathematical Physics}}
  \bibinfo{volume}{50}(\bibinfo{number}{12}), p. \bibinfo{pages}{122106},
  \doi{10.1063/1.3271109}.

\bibitemdeclare{article}{Banuls}
\bibitem{Banuls}
\bibinfo{author}{Mari-Carmen \surnamestart Ba\~{n}uls\surnameend},
  \bibinfo{author}{C.~\surnamestart Navarrete\surnameend},
  \bibinfo{author}{A.~\surnamestart P\'erez\surnameend},
  \bibinfo{author}{Eugenio \surnamestart Rold\'{a}n\surnameend} \&
  \bibinfo{author}{J.~\surnamestart Soriano\surnameend} (\bibinfo{year}{2006}):
  \emph{\bibinfo{title}{Quantum Walk with a time-dependent coin}}.
\newblock {\sl \bibinfo{journal}{Physical Review A}}
  \bibinfo{volume}{73}(\bibinfo{number}{6}), \doi{10.1103/PhysRevA.73.062304}.

\bibitemdeclare{article}{Cedzich2016}
\bibitem{Cedzich2016}
\bibinfo{author}{C.~\surnamestart Cedzich\surnameend} \& \bibinfo{author}{R.~F.
  \surnamestart Werner\surnameend} (\bibinfo{year}{2016}):
  \emph{\bibinfo{title}{Revivals in quantum walks with a
  quasiperiodically-time-dependent coin}}.
\newblock {\sl \bibinfo{journal}{Physical Review A}}
  \bibinfo{volume}{93}(\bibinfo{number}{3}), \doi{10.1103/physreva.93.032329}.

\bibitemdeclare{article}{Chen}
\bibitem{Chen}
\bibinfo{author}{Tian \surnamestart Chen\surnameend},
  \bibinfo{author}{Bo~\surnamestart Wang\surnameend} \&
  \bibinfo{author}{Xiangdong \surnamestart Zhang\surnameend}
  (\bibinfo{year}{2017}): \emph{\bibinfo{title}{Controlling probability
  transfer in the discrete-time quantum walk by modulating the symmetries}}.
\newblock {\sl \bibinfo{journal}{New Journal of Physics}}
  \bibinfo{volume}{19}(\bibinfo{number}{11}), p. \bibinfo{pages}{113049},
  \doi{10.1088/1367-2630/aa8fe4}.

\bibitemdeclare{article}{Childs2009}
\bibitem{Childs2009}
\bibinfo{author}{Andrew~M. \surnamestart Childs\surnameend}
  (\bibinfo{year}{2009}): \emph{\bibinfo{title}{Universal Computation by
  Quantum Walk}}.
\newblock {\sl \bibinfo{journal}{Phys. Rev. Lett.}} \bibinfo{volume}{102}, p.
  \bibinfo{pages}{180501}, \doi{10.1103/PhysRevLett.102.180501}.

\bibitemdeclare{article}{DiMolfetta2015}
\bibitem{DiMolfetta2015}
\bibinfo{author}{Giuseppe \surnamestart Di~Molfetta\surnameend},
  \bibinfo{author}{Lauchlan \surnamestart Honter\surnameend},
  \bibinfo{author}{Ben~B. \surnamestart Luo\surnameend},
  \bibinfo{author}{Tatsuaki \surnamestart Wada\surnameend} \&
  \bibinfo{author}{Yutaka \surnamestart Shikano\surnameend}
  (\bibinfo{year}{2015}): \emph{\bibinfo{title}{Massless Dirac equation from
  Fibonacci discrete-time quantum walk}}.
\newblock {\sl \bibinfo{journal}{Quantum Studies: Mathematics and Foundations}}
  \bibinfo{volume}{2}(\bibinfo{number}{3}), p. \bibinfo{pages}{243},
  \doi{10.1007/s40509-015-0038-6}.

\bibitemdeclare{article}{Suzanne}
\bibitem{Suzanne}
\bibinfo{author}{Suzanne \surnamestart Greist-Bousquet\surnameend} \&
  \bibinfo{author}{Harvey~R \surnamestart Schiffman\surnameend}
  (\bibinfo{year}{1981}): \emph{\bibinfo{title}{The Poggendorff Illusion: An
  Illusion of Linear Extent?}}
\newblock {\sl \bibinfo{journal}{Perception}}
  \bibinfo{volume}{10}(\bibinfo{number}{2}), pp. \bibinfo{pages}{155--164},
  \doi{10.1068/p100155}.
\newblock \bibinfo{note}{PMID: 7279544}.

\bibitemdeclare{article}{Karski174}
\bibitem{Karski174}
\bibinfo{author}{Michal \surnamestart Karski\surnameend},
  \bibinfo{author}{Leonid \surnamestart F{\"o}rster\surnameend},
  \bibinfo{author}{Jai-Min \surnamestart Choi\surnameend},
  \bibinfo{author}{Andreas \surnamestart Steffen\surnameend},
  \bibinfo{author}{Wolfgang \surnamestart Alt\surnameend},
  \bibinfo{author}{Dieter \surnamestart Meschede\surnameend} \&
  \bibinfo{author}{Artur \surnamestart Widera\surnameend}
  (\bibinfo{year}{2009}): \emph{\bibinfo{title}{Quantum Walk in Position Space
  with Single Optically Trapped Atoms}}.
\newblock {\sl \bibinfo{journal}{Science}}
  \bibinfo{volume}{325}(\bibinfo{number}{5937}), pp. \bibinfo{pages}{174--177},
  \doi{10.1126/science.1174436}.

\bibitemdeclare{article}{Knight}
\bibitem{Knight}
\bibinfo{author}{Peter~L. \surnamestart Knight\surnameend},
  \bibinfo{author}{Eugenio \surnamestart Rold\'{a}n\surnameend} \&
  \bibinfo{author}{J.~E. \surnamestart Sipe\surnameend} (\bibinfo{year}{2004}):
  \emph{\bibinfo{title}{Propagating quantum walks: The origin of interference
  structures}}.
\newblock {\sl \bibinfo{journal}{Journal of Modern Optics}}
  \bibinfo{volume}{51}(\bibinfo{number}{12}), p. \bibinfo{pages}{1761},
  \doi{10.1080/09500340408232489}.

\bibitemdeclare{article}{Lovett2010}
\bibitem{Lovett2010}
\bibinfo{author}{Neil~B. \surnamestart Lovett\surnameend},
  \bibinfo{author}{Sally \surnamestart Cooper\surnameend},
  \bibinfo{author}{Matthew \surnamestart Everitt\surnameend},
  \bibinfo{author}{Matthew \surnamestart Trevers\surnameend} \&
  \bibinfo{author}{Viv \surnamestart Kendon\surnameend} (\bibinfo{year}{2010}):
  \emph{\bibinfo{title}{Universal quantum computation using the discrete-time
  quantum walk}}.
\newblock {\sl \bibinfo{journal}{Physical Review A}}
  \bibinfo{volume}{81}(\bibinfo{number}{4}), \doi{10.1103/physreva.81.042330}.

\bibitemdeclare{article}{Panahiyan2018}
\bibitem{Panahiyan2018}
\bibinfo{author}{S~\surnamestart Panahiyan\surnameend} \&
  \bibinfo{author}{S~\surnamestart Fritzsche\surnameend}
  (\bibinfo{year}{2018}): \emph{\bibinfo{title}{Controlling quantum random walk
  with a step-dependent coin}}.
\newblock {\sl \bibinfo{journal}{New Journal of Physics}}
  \bibinfo{volume}{20}(\bibinfo{number}{8}), p. \bibinfo{pages}{083028},
  \doi{10.1088/1367-2630/aad899}.

\bibitemdeclare{article}{Shikano2014}
\bibitem{Shikano2014}
\bibinfo{author}{Y.~\surnamestart Shikano\surnameend},
  \bibinfo{author}{T.~\surnamestart Wada\surnameend} \&
  \bibinfo{author}{J.~\surnamestart Horikawa\surnameend}
  (\bibinfo{year}{2014}): \emph{\bibinfo{title}{Discrete-time quantum walk with
  feed-forward quantum coin}}.
\newblock {\sl \bibinfo{journal}{Scientific Reports}} \bibinfo{volume}{4},
  \doi{10.1038/srep04427}.

\bibitemdeclare{article}{wojcik2004quasiperiodic}
\bibitem{wojcik2004quasiperiodic}
\bibinfo{author}{Antoni \surnamestart W{\'o}jcik\surnameend},
  \bibinfo{author}{Tomasz \surnamestart {\L}uczak\surnameend},
  \bibinfo{author}{Pawe{\l} \surnamestart Kurzy{\'n}ski\surnameend},
  \bibinfo{author}{Andrzej \surnamestart Grudka\surnameend} \&
  \bibinfo{author}{Ma{\l}gorzata \surnamestart Bednarska\surnameend}
  (\bibinfo{year}{2004}): \emph{\bibinfo{title}{Quasiperiodic dynamics of a
  quantum walk on the line}}.
\newblock {\sl \bibinfo{journal}{Physical review letters}}
  \bibinfo{volume}{93}(\bibinfo{number}{18}), p. \bibinfo{pages}{180601},
  \doi{10.1038/nature02008}.

\bibitemdeclare{article}{Xue}
\bibitem{Xue}
\bibinfo{author}{P.~\surnamestart Xue\surnameend},
  \bibinfo{author}{R.~\surnamestart Zhang\surnameend},
  \bibinfo{author}{H.~\surnamestart Qin\surnameend},
  \bibinfo{author}{X.~\surnamestart Zhan\surnameend}, \bibinfo{author}{Z.~H.
  \surnamestart Bian\surnameend}, \bibinfo{author}{J.~\surnamestart
  Li\surnameend} \& \bibinfo{author}{Barry~C. \surnamestart Sanders\surnameend}
  (\bibinfo{year}{2015}): \emph{\bibinfo{title}{Experimental Quantum-Walk
  Revival with a Time-Dependent Coin}}.
\newblock {\sl \bibinfo{journal}{Phys. Rev. Lett.}} \bibinfo{volume}{114}, p.
  \bibinfo{pages}{140502}, \doi{10.1103/PhysRevLett.114.140502}.

\bibitemdeclare{article}{Yalnkaya}
\bibitem{Yalnkaya}
\bibinfo{author}{\.{I}skender \surnamestart Yal\c{c}{\i}nkaya\surnameend} \&
  \bibinfo{author}{Zafer \surnamestart Gedik\surnameend}
  (\bibinfo{year}{2015}): \emph{\bibinfo{title}{Qubit state transfer via
  discrete-time quantum walks}}.
\newblock {\sl \bibinfo{journal}{Journal of Physics A: Mathematical and
  Theoretical}} \bibinfo{volume}{48}(\bibinfo{number}{22}), p.
  \bibinfo{pages}{225302}, \doi{10.1088/1751-8113/48/22/225302}.

\bibitemdeclare{article}{Zhan2014}
\bibitem{Zhan2014}
\bibinfo{author}{Xiang \surnamestart Zhan\surnameend}, \bibinfo{author}{Hao
  \surnamestart Qin\surnameend}, \bibinfo{author}{Zhi-hao \surnamestart
  Bian\surnameend}, \bibinfo{author}{Jian \surnamestart Li\surnameend} \&
  \bibinfo{author}{Peng \surnamestart Xue\surnameend} (\bibinfo{year}{2014}):
  \emph{\bibinfo{title}{Perfect state transfer and efficient quantum routing: A
  discrete-time quantum-walk approach}}.
\newblock {\sl \bibinfo{journal}{Phys. Rev. A}} \bibinfo{volume}{90}, p.
  \bibinfo{pages}{012331}, \doi{10.1103/PhysRevA.90.012331}.

\end{thebibliography}
\end{document}